\documentclass[main]{aa}  
\usepackage{graphicx}
\usepackage{txfonts}
\usepackage{amsmath}
\def\h{\hskip -0.3 mm}
\def\oiii{[O\,{\sc iii}]}

\bibpunct{(}{)}{;}{a}{}{,} 

\begin{document}

\title{The Size-luminosity Relation of the AGN Torus Determined from the Comparison between Optical and Mid-infrared Variability}

\author{Minjin Kim\inst{1} \and Suyeon Son\inst{1} \and Luis C. Ho\inst{2,3}}

\institute{Department of Astronomy and Atmospheric Sciences,
Kyungpook National University, Daegu 41566, Korea\\
              \email{mkim.astro@gmail.com}
         \and
             Kavli Institute for Astronomy and Astrophysics, Peking University, Beijing 100871, China
         \and
             Department of Astronomy, School of Physics, Peking University, Beijing 100871, China
             }

\date{Received}

\abstract{
We investigate the optical variability of low-redshift ($0.15< z\leq0.4$) active galactic nuclei using the multi-epoch data from the Zwicky Transient Facility. We find that a damped random walk model well describes the ensemble structure function in the $g$ band. Consistent with previous studies, more luminous active galactic nuclei tend to have a steeper structure function at a timescale less than the break timescale and smaller variability amplitude. By comparing the structure functions in the optical with the mid-infrared obtained from the Wide-field Infrared Survey Explorer, we derive the size of the dusty torus using a toy model for the geometry of the torus. The size of the torus positively correlates with the luminosity of the active nucleus, following a relation that agrees well with previous studies based on reverberation mapping. This result demonstrates that the structure function method can be used as a powerful and highly efficient tool to examine the size of the torus.}        


\titlerunning{Torus $R-L$ Relation}
\authorrunning{Kim et al.}

\maketitle

\section{Introduction}

The ultraviolet-optical continuum from unobscured active galactic nuclei (AGNs) originates from the thermal emission from the accretion disk. It is known to vary over a wide range of timescales from days to years \cite[e.g.,][]{urlich_1997}. While the physical origin of the ultraviolet-optical variability is still under debate, recent observational studies argue that it could be attributed to thermal instability, in which the variability timescale scales with the orbital timescale or black hole mass \cite[e.g.,][]{kelly_2009,sun_2020,burke_2021,arevalo_2023, tang_2023}. The intrinsic variability is often modeled with a damped random walk, in which the power spectral density (PSD) decreases with increasing frequency as a power law (i.e., PSD $\propto \nu^{-2}$) at high frequency above the break frequency ($\nu \gg \nu_{\rm break}$) and flattens at low frequency ($\nu \leq \nu_{\rm break}$; \citealt{kelly_2011}). At the same time, AGNs exhibit a strong mid-infrared (MIR) continuum emitted from dust heated in a donut-shaped torus around the supermassive black hole and accretion disk. MIR emission also varies in response to the ultraviolet-optical continuum \cite[e.g.,][]{kozlowski_2016a,son_2022}. The characteristics of MIR variability are mostly determined by the combination of the intrinsic variability in the accretion disk and the geometry of the dusty torus, indicating that it can be used to probe the structure of the torus \cite[e.g.,][]{li_2023, son_2023b}.   

Variability has been widely used to investigate the physical properties of the central structures in AGNs. In particular, the reverberation mapping (RM; \citealt{blandford_1982}) technique allows us to derive the size of central components robustly. This method has been applied primarily to estimate the size of the broad-line region (BLR) by estimating the time lag between the ultraviolet-optical continuum from the accretion disk and the fluxes of various broad emission lines emitted from the dense material in the BLR, which is ionized by the high-energy photons from the accretion disk \cite[e.g.,][]{peterson_2004, bentz_2009, du_2015, shen_2023, woo_2024}. Recently, the size of the accretion disk has been evaluated for a handful of nearby AGNs (e.g., Mrk 142, NGC 4593, NGC 5548) using X-ray and ultraviolet-optical continuum reverberation mapping \cite[e.g.,][]{fausnaugh_2016, cackett_2018, cackett_2020}. 

The time lag between the ultraviolet-optical continuum and IR continuum has been used to constrain the size of the torus (\citealp{suganuma_2006,koshida_2014} for $K$-band, and \citealp{lyu_2019,yang_2020,mandal_2024} for $3.4-4.6\ \mu$m), with the aid of multi-epoch near-infrared (NIR) data from ground-based telescopes and MIR light curves from the Wide-field Infrared Survey Explorer (WISE; \citealp{wright_2010}). These studies reveal that the size of the torus is positively correlated with the AGN luminosity. Prior or parallel to these RM works, observational studies with the MIR interferometry also demonstrated similar results \cite[e.g.][]{kishimoto_2011,burtscher_2013, gravity_2020b}. This size-luminosity relation is also found in the BLR reverberation mapping studies. In addition, the size of the torus increases with increasing rest-frame wavelength, indicating the inner part of the torus is hotter than the outer part, consistent with the torus model \cite[e.g.,][]{lyu_2019}. However, because the reverberation mapping method requires intensive observations of high cadence and long duration for individual objects, it can only be applied to a limited number of targets. 

Alternatively, \citet{li_2023} demonstrated that the comparison between the ultraviolet-optical and MIR variability can provide statistically valuable constraints on the size of the dusty torus. Although this method cannot be applied to an individual object, it permits the investigation of the average geometry of the torus in a sample of AGNs sharing similar properties. Moreover, unlike the reverberation mapping method, it does not require simultaneous, multi-epoch, multiwavelength data with a regular cadence for a single target. Therefore, this method is suitable for existing time-series surveys.       

AGN variability is known to follow a damped random walk, which is well described by a broken power-law PSD (e.g., \citealp{kelly_2009, Kozlowski_2010}; but see \citealp{mushotzky_2011, kasliwal_2015}). However, characterization of the PSD requires multi-epoch data with regular and dense sampling, as a consequence of which the PSD of ultraviolet-optical light curves has been estimated robustly for relatively small samples \citep{smith_2018}. An alternative, effective strategy to study the characteristics of AGN variability utilizes the structure function (SF), which is defined as the root-mean-square of the magnitudes at a given time difference $\Delta t$. The SF is relatively free from bias due to sparse and irregular sampling, and it reasonably approximates a damped random walk \cite[e.g.,][]{hughes_1992, macleod_2010, kozlowski_2016b}. On the other hand, unless the multi-epoch data are well sampled with a dense cadence and sufficiently long baseline, it is hard to fully constrain the SF of a single object \citep{kozlowski_2016b}. To overcome this limitation, one can utilize an ensemble SF computed by averaging the SFs from multiple objects at a given timescale. In particular, ensemble SFs have been widely used to examine how AGN variability depends on the physical properties of AGNs \cite[e.g.,][]{trevese_1994, vandenberk_2004, son_2023b}.  

\citet{li_2023} adopted a simple toy model for the torus, one solely determined by the torus geometry without considering radiative transfer, in order to compute the torus transfer function. From a direct comparison between the ensemble SFs of the optical and MIR variability, in conjunction with the torus transfer function, \citet{li_2023} investigated the size of the torus for a sample of relatively luminous [$\log~ (L_{\rm bol}/{\rm erg~s^{-1}}) \ge 45.3$] and distant ($0.5\le z \le 1.2$) quasars selected from the multi-epoch optical photometric data from Stripe 82 of the Sloan Digital Sky Survey (SDSS; \citealp{abazajian_2009, annis_2014}) and MIR photometry from WISE. In good agreement with previous studies conducted with the reverberation mapping method, this experiment showed that the size of the torus positively correlates with the AGN luminosity \citep{li_2023}.

In this study, we apply the methodology of \citet{li_2023} to examine whether the size-luminosity relation applies to quasars of lower redshift ($0.15< z \le 0.4$) and lower luminosity [$44.4 \le \log~(L_{\rm bol}/{\rm erg~s^{-1}}) \le 45.9$]. In particular, the narrow range of redshift is essential to minimize the dependence of the torus size on rest-frame wavelength. Section 2 describes the sample selection and data. Section 3 presents the detailed method to construct the ensemble SFs. The comparison between the optical and MIR SFs is shown in Section 4, and size-luminosity relation from our method is given in Section 5. Section 6 summarizes the overall results. We adopt the cosmological parameters $H_0=100h=67.4$ km ${\rm s}^{-1}$ ${\rm Mpc}^{-1}$, $\Omega_m=0.315$, and $\Omega_\Lambda=0.685$ (\citealt{planck_2020}).  

\begin{figure}[tbp!]
\centering
\includegraphics[width=0.45\textwidth]{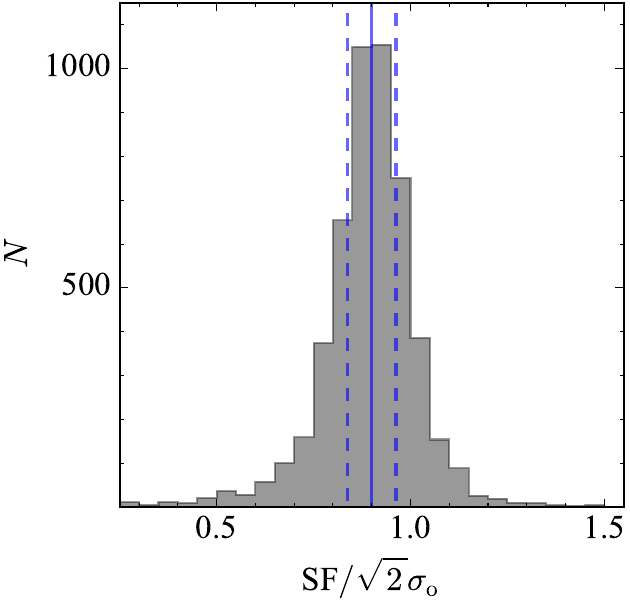}
\caption{
Distribution of SFs divided by $\sqrt{2}$ and the original photometric uncertainty ($\sigma_o$) of the $g$-band light curves of the non-variable sources. The blue solid line denote the median value ($0.90$), and the dashed lines give its median absolute deviation ($0.06$). If the photometric noise is estimated robustly, the median value should be equal to 1; the median value of 0.90 indicates that the photometric uncertainty is underestimated.
}
\end{figure}

\section{Sample and data}

For a direct comparison between the optical and MIR SFs, we adopt the same sample of type~1 AGNs from \citet{son_2023b}, for which the MIR SF was computed using the light curves from WISE. The parent sample was initially drawn from the Data Release 14 quasar catalog of the SDSS (\citealp{paris_2018}). To minimize the effect of cosmic evolution on the dust properties and photometric uncertainties introduced by the extended features of the host galaxy, \citet{son_2023b} imposed a redshift cut of $0.15<z\le0.4$. To secure NIR and MIR data, \citet{son_2023b} selected sources with 2MASS (\citealp{skrutskie_2006}) and WISE counterparts with a matching radius of 2$^{\prime\prime}$, to arrive at a sample of 4,295 type 1 AGNs. \citet{son_2023b} demonstrated that the NIR data are crucial to constrain properly the flux contribution from the host galaxy \citep{son_2022,son_2023b}.

To obtain the optical light curves in the $g$ band, we crossmatch the initial sample from \citet{son_2023b} with the Zwicky Transient Facility (ZTF) Data Release 19 using a matching radius of 2$^{\prime\prime}$ \citep{bellm_2019}. We find that all the sources have counterparts in ZTF. The average baseline and cadence of the optical light curves are $\sim1785$ days and $\sim21.7$ days, respectively. To remove any suspicious measurements, we only use the photometric data with flag=0, separation angle $< 0\farcs4$, and limiting $g$-band magnitude $< 20.2$. These criteria were carefully determined from visual inspection of the light curves to discard outliers in the photometric measurements (e.g., timescale of spikes and dips less than a few days). To reconstruct the SF robustly, we choose sources that have ZTF photometric data from at least 20 epochs. Finally, to only consider variable sources, we adopt the criterion $P_{\rm var}\ge0.95$, which is the probability that the target is genuinely variable, calculated from $\chi^2$ statistics \citep{sanchez_2018, son_2023b}. $P_{\rm var}$ is estimated after the host galaxy contribution is subtracted. Note that the selection criterion of 0.95 is suitable for selecting weakly or moderately variable sources: $\sim87$\% of the entire sample meets this criterion. Without this selection, SFs with large photometric uncertainties can introduce systematic bias in the ensemble estimation. Radio-loud AGNs may exhibit enhanced variability due to the relativistic jet. We compute the radio-loudness of the sample using the 1.4GHz luminosity ($L_{\rm 1.4 GHz}$) from the FIRST survey and $i$-band luminosity ($L_{i}$), in which a $K$-correction is applied by assuming a spectral index of $-0.5$ \cite[][]{richards_2006, kimball_2008}. The radio-loud AGNs are classified with a criterion of $R>1$, where $R$ is defined as $L_{\rm 1.4 GHz}/L_i$ \citep{balokovic_2012}, finally excluded from further analysis. This reduces the final sample to 3,578 objects. 

The bolometric luminosity is estimated from the continuum luminosity at 5100\,\AA\ with a bolometric correction of $L_{\rm bol}=9.26\,L_{5100}$ derived from the mean spectral energy distribution of type 1 quasars (\citealp{richards_2006}), where $L_{5100}$ is computed from spectral fitting of the SDSS spectra (\citealp{rakshit_2020}). We adopt the bolometric luminosities based on 
$L_{5100}$ instead of alternatives such those derived from the strength of \oiii $\lambda 5007$ \citep{son_2023b}, which is sensitive to the covering factor of the narrow-line region, the ionization parameter, and the shape of the ionizing continuum \cite[e.g.,][]{heckman_2004,shen_2011,kong_2018,netzer_2019}. While the bolometric correction is known to be sensitive to the AGN luminosity, its uncertainty is typically smaller than 0.1 dex in the range of the AGN luminosity in our sample \cite[e.g.,][]{hopkins_2007, netzer_2019}.  
The final sample has a median $z = 0.296$, $M_g = -22.03$ mag, $\log~(M_{\rm BH}/M_\odot) = 8.19$, $\log\ (L_{\rm bol}/{\rm erg~s^{-1}}) = 45.2$, and $\log~(L_{\rm bol}/L_{\rm Edd}) = -1.05$. We follow \citet{son_2023b} and calculate the black hole mass $M_{\rm BH}$ using the virial method and spectral measurements of \citet{rakshit_2020}.

\begin{figure}[tbp!]
\centering
\includegraphics[width=0.45\textwidth]{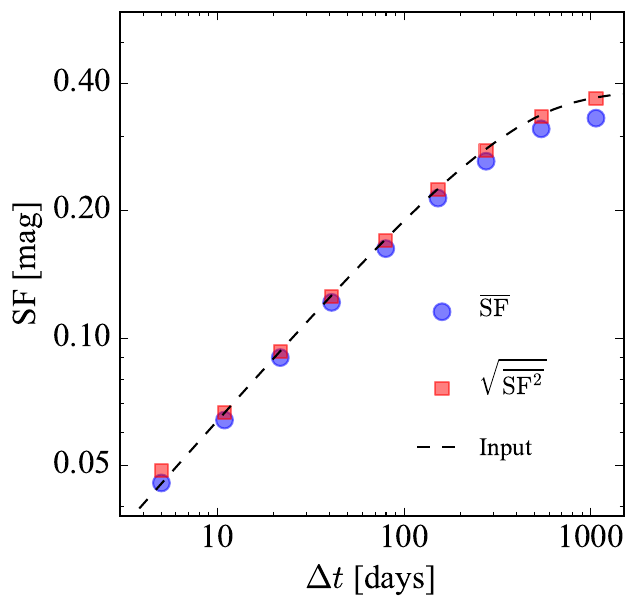}
\caption{
Ensemble SFs derived from the mock light curves. The 10,000 light curves were generated to mimic the real multi-epoch data from ZTF. The dashed line denotes the input SF. The red squares and blue circles represent the square root of the mean ${\rm SF}^2$ and the mean of SF, respectively.
}
\end{figure}

\section{Structure function}
We quantify the optical variability with the SF, defined as the mean difference of the magnitude in the light curve as a function of the time lag ($\Delta t$):
\begin{equation}
\begin{split}
    {\rm SF}^2(\Delta t) = \frac{1}{N_{\Delta t, {\rm pair}}} \sum_{i=1}^{N_{\Delta t, {\rm pair}}} (m(t) - m(t+\Delta t))^2 \\ - \sigma_{\rm e}^2(t) - \sigma_{\rm e}^2(t+\Delta t), 
\end{split}
\end{equation}
\noindent
where $N_{\Delta t, {\rm pair}}$ denotes the number of pairs associated with $\Delta t$, $m$ is the magnitude, and $\sigma_{\rm e}$ represents the uncertainty of the magnitude in each epoch. In general, AGN light curves can be described by a damped random walk in which the SF is proportional to the time lag at $\Delta t < \tau$ (${\rm SF} \propto \Delta t^1$) and flattens at $\Delta t \ge \tau$ (${\rm SF} \propto \Delta t^0$), where $tau$ is the break timescale. This can be expressed more generally as 

\begin{equation}
{\rm SF} = [{\rm SF^2_{\infty}} (1-e^{(-\Delta t/\tau)^\gamma}) ]^{0.5},
\end{equation} 

\noindent
where ${\rm SF_{\infty}}$ is the SF at $\Delta t \gg \tau$ and $\gamma$ is the power-law slope for $\Delta t < \tau$ (i.e., ${\rm SF} \propto \Delta t^\gamma$). The damped random walk model can be expressed with $\gamma=1$. The uncertainty of the SF for an individual target is computed from the square root of the standard deviation of $\Delta {\rm mag} \equiv (m(t) - m(t+\Delta t))^2 - \sigma_{\rm e}^2(t) - \sigma_{\rm e}^2(t+\Delta t)$ at a given $\Delta t$ (see Equation 1).

\subsection{Host galaxy contribution}
Although the ZTF survey provides photometry in both $g$ and $r$, we use the $g$-band data because it is less contaminated by the flux from the host galaxy. Nevertheless, we estimate the flux contribution from the host in the $g$ band by fitting the spectral energy distribution that spans from the optical to the MIR, constructed from measurements from SDSS, 2MASS, and WISE. Following the detailed fitting method described in \citet{son_2023}, we adopt three spectral energy distribution templates for AGNs (hot dust-deficient, warm dust-deficient, and normal AGNs) from \citet{lyu_2017a} and seven templates for the host galaxy from \citet{polletta_2007}, which comprise an old (7~Gyr) stellar population and galaxies of Hubble type E, S0, Sa, Sb, Sc, and Sd. The derived host flux from the individual target is subtracted from the original light curve. The average host fraction in the $g$ band is only $\sim0.12$, which has a minor effect on the SF measurements.

\subsection{Photometric uncertainty}
The SF is known to be highly sensitive to photometric error, and its effect is severe when estimating the slope on short timescales \cite[e.g.][]{kozlowski_2016b}. To check whether the photometric uncertainties ($\sigma_{\rm o}$) provided by the ZTF survey are reliable, we use the ZTF light curves of non-variable sources drawn from the inactive galaxies from the Max Planck Institute for Astrophysics and the Johns Hopkins University (MPA–JHU) catalog \citep{brinchmann_2004}. From the SF calculated from the individual object without removing the contribution from the photometric noise, ${\rm SF}^2(\Delta t) = \frac{1}{N_{\Delta t, {\rm pair}}} \sum_{i=1}^{N_{\Delta t, {\rm pair}}} (m(t) - m(t+\Delta t))^2$. We apply the same criteria as the science data for the AGN sample to select reliable photometric measurements. If the original photometric error ($\sigma_{\rm o}$) is estimated robustly, the SFs are expected to equal $\sqrt{2} \sigma_{\rm o}$. Instead, we find ${\rm SF} / \sqrt{2} \approx 0.9\sigma_{\rm o}$ (Figure~1), which indicates that the photometric noise is underestimated. To account for this, we adopt a final photometric error of $\sigma_{\rm e} = 0.9 \sigma_{\rm o}$.

\begin{table*}
\centering
\caption{Structure function parameters\label{tab:table1}}
\begin{tabular}{c c c c c c c c}
\hline \hline
Subsample &
${\rm SF_{\infty}}$ &
$\gamma$ &
$\tau$ (days) &
 &
${\rm SF_{\infty}}$ &
$\gamma$ (fixed) &
$\tau$ (days) \\
\cline{2-4} \cline{6-8}
(1) &
(2) &
(3) &
(4) &
 &
(6) &
(7) &
(8) \\
\hline
All                              & $0.37\pm0.02$ & $0.99\pm0.05$ & $326\pm54$ && $0.37\pm0.02$ & $1$ & $314\pm50$\\
$\log L_{\rm bol} \leq 44.67$    & $0.40\pm0.02$ & $0.81\pm0.06$ & $281\pm83$ && $0.36\pm0.02$ & $1$ & $155\pm36$\\
$44.67<\log L_{\rm bol}\leq45.17$& $0.38\pm0.01$ & $0.98\pm0.04$ & $278\pm42$ && $0.38\pm0.01$ & $1$ & $265\pm38$\\
$45.17<\log L_{\rm bol}\leq45.67$& $0.36\pm0.02$ & $1.14\pm0.05$ & $293\pm44$ && $0.40\pm0.05$ & $1$ & $482\pm190$\\
$45.67<\log L_{\rm bol}$         & $0.27\pm0.01$ & $1.26\pm0.03$ & $277\pm22$ && $0.35\pm0.11$ & $1$ & $751\pm658$\\
\hline
\end{tabular}
\tablefoot{Col. (1): Subsample, where $L_{\rm bol}$ is the AGN bolometric luminosity in units of erg s$^{-1}$.
Col. (2): SF amplitude at $\Delta t \gg \tau$.
Col. (3): Power-law index at $\Delta t < \Delta \tau$.
Col. (4): Break in the power law.
Col. (5): SF amplitude at $\Delta t \gg \tau$ with a fixed $\gamma=1$.
Col. (6): Power-law index at $\Delta t < \Delta \tau$ fixed as 1.
Col. (7): Break in the power law with a fixed $\gamma=1$.
}
\end{table*}

\begin{figure*}[tbp!]
\centering
\includegraphics[width=8.5cm]{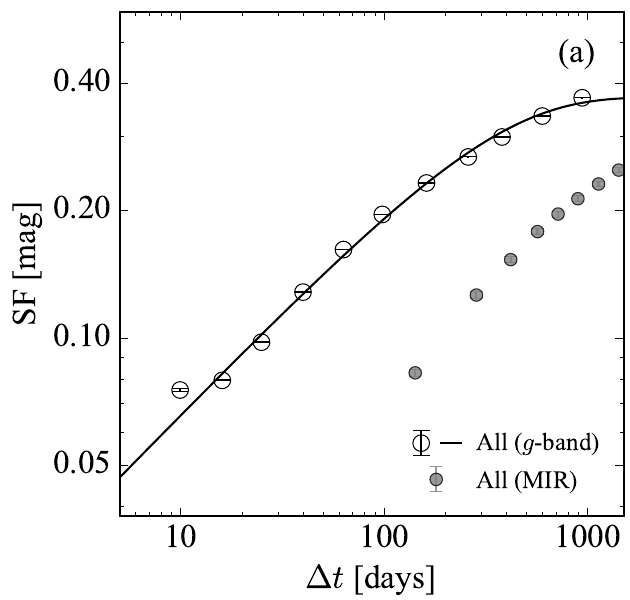}
\includegraphics[width=8.5cm]{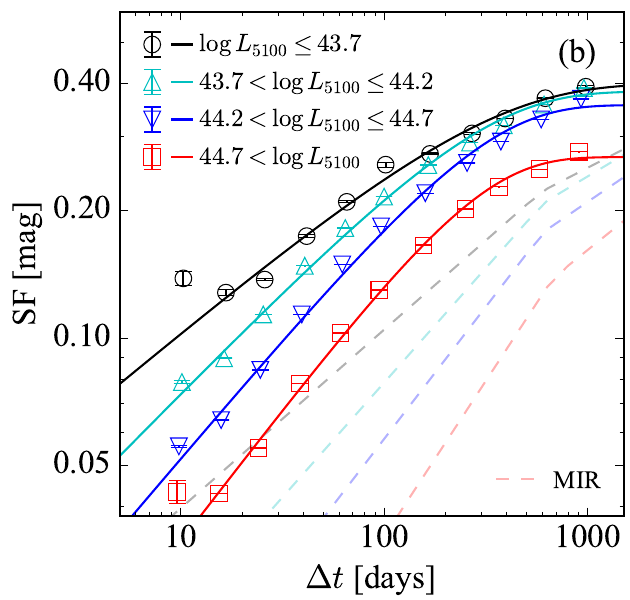}
\caption{
(a) Ensemble SFs of the entire sample. Open circles represent the SF for the optical data from ZTF, with the solid line denoting their fit to a broken power law. Filled circles give the SF for the MIR data from \citet{son_2023b}. (b) Ensemble SFs of the subsamples binned by the continuum luminosity at 5100\,\AA. Open symbols and solid lines are the SFs and fitting results for the optical data, while the dashed lines show the corresponding SFs for the MIR data.
}
\end{figure*}

\subsection{Ensemble structure function}
As in the work of \citet{son_2023b} on the MIR SF of AGNs, this study utilizes the ensemble SF, which is the average value of the SFs for a number of objects at a given $\Delta t$. The ensemble SFs are useful to examine statistically the characteristics of the variability as a function of various AGN properties, using light curves with sparse and irregular cadences \cite[e.g.,][]{vandenberk_2004,bauer_2009}. 

The reliability of the SF also depends on the cadence and baseline of the light curves. To test this, we made 10,000 mock light curves modeled with a damped random walk, for which the sampling of the mock light curve is randomly taken from the real observations of ZTF for our sample. To mimic the ZTF light curves of our sample, we generate the input light curves assuming that ${\rm SF_\infty} = 0.38$, $\tau=350$ days, $\gamma=1$, and that the photometric uncertainty is 0.068 mag. We then compute the ensemble SF using two different estimators: the square root of the mean ${\rm SF}^2$ and the mean of SF. Our simulations show that the ensemble SF is slightly overestimated at $\Delta t < 15~{\rm days}$, as is also seen in the observed data (Figure 2). If the photometric error is larger than the intrinsic SF, the mean observed ${\rm SF}^2$ occasionally could be smaller than zero, which should be excluded in the calculation of the ensemble SF. This can boost the SF at small $\Delta t$, where the intrinsic SF is smaller than the photometric noise. Therefore, this study uses the ensemble SF at $\Delta t \ge 15~{\rm days}$. \citet{son_2023b} demonstrated, and the simulation in Figure 2 confirms, that the square root of the mean ${\rm SF}^2$ is a better tracer of the real ensemble SF than the mean of SF. We also compute the ensemble SFs in the same manner. Finally, the uncertainties in the ensemble SF are estimated from bootstrap resampling of the SFs of the individual target within the uncertainty of the individual SF. We perform 1000 realizations of resampling, and the final uncertainty is estimated from half of the difference between the 16th and 84th percentiles.

\begin{figure*}[t!]
\centering
\includegraphics[width=6cm]{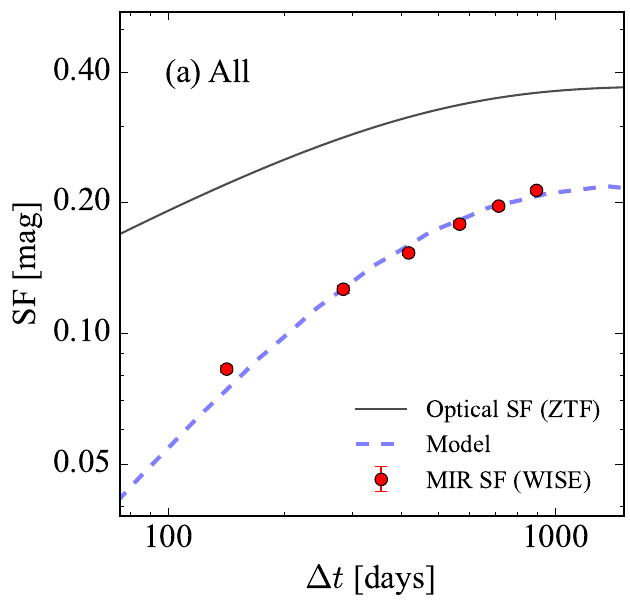}
\includegraphics[width=6cm]{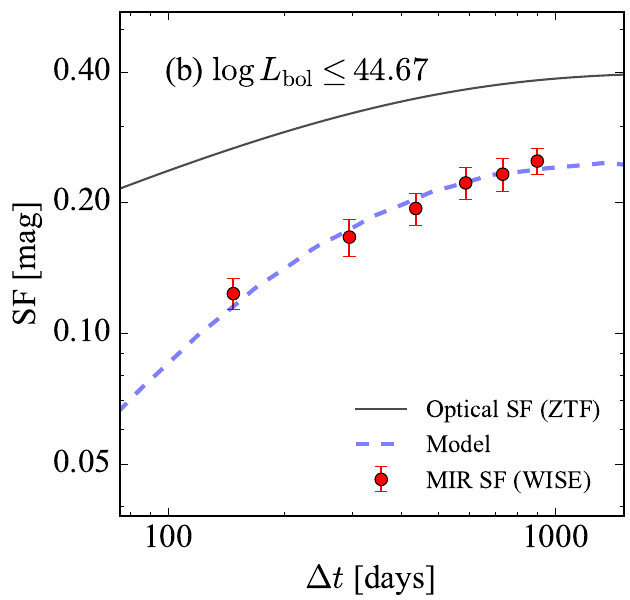}
\includegraphics[width=6cm]{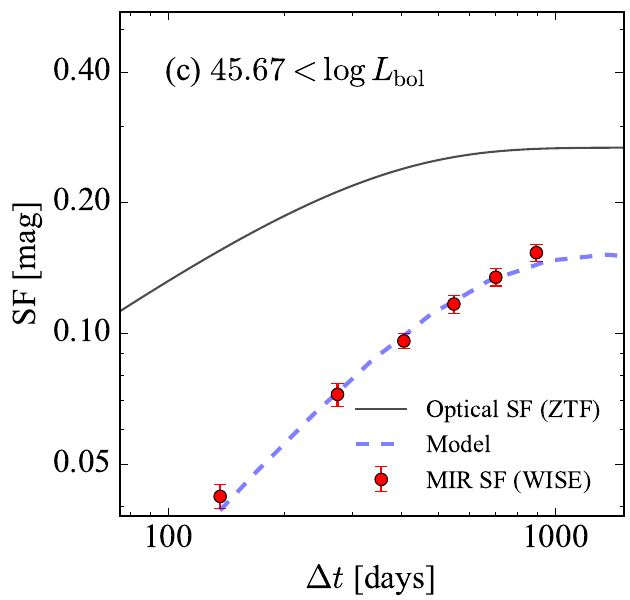}
\caption{
Fitting results of the MIR SFs for (a) the entire sample, (b) the least luminous sources, and (c) the most luminous sources. The solid line is the optical SF estimated from the ZTF survey. Red points represent the MIR SF from the WISE survey, with the blue dashed line denoting the best fit.
}
\end{figure*}

\section{Results}
\subsection{The $g$-band structure function}

We first estimate the ensemble $g$-band SF of the entire sample (Figure 3; Table 1). The break timescale $\tau$ is known to be correlated with black hole mass and AGN luminosity \cite[e.g.,][]{kelly_2009,macleod_2010,burke_2021,tang_2023}. However, the relatively short time baseline of the ZTF survey makes it difficult to estimate $\tau$ robustly with this dataset. Although we only use Equation (2) to investigate the qualitative form of the SF, without attaching any specific physical meaning to it at this stage, we nevertheless derive a power-law slope of $\gamma=0.99\pm0.05$, which agrees well with the prediction from the damp random walk model. 

Dividing the sample into four luminosity bins reveals that $\gamma$ is proportional to luminosity, such that more luminous AGNs exhibit a steeper SF over short timescales than fainter sources. The parameter ${\rm SF_{\infty}}$ is inversely correlated with the AGN luminosity (Table 1). These findings are consistent with the trends reported in previous studies, not only in terms of MIR variability \citep{son_2023b} but also optical variability \cite[e.g.,][]{vandenberk_2004,macleod_2010,tang_2023}.

\subsection{Fitting with torus model}
The MIR light curves can be expressed as the combination of the intrinsic light curve from the accretion disk and the transfer function of the dusty torus,

\begin{equation}
L_{\rm MIR}(t) = \int{\Psi(\tau)L_{\rm opt}(t-\tau) d\tau},
\end{equation}

\noindent
where $L_{\rm MIR}$ is the MIR light curve, $\Psi(\tau)$ is the transfer function of the torus, and $L_{\rm opt}$ is the ultraviolet-optical continuum light curve. We adopt the torus model of \citet{li_2023}, for which the transfer function of the torus is determined solely by its geometry, which is described by an inner radius ($R_{\rm in}$), a half-opening angle ($\theta$) representing the angle between the equatorial plane and the edge of the torus, an outer to inner radius ratio ($Y$), and an inclination angle ($i$). A total of 50,000 clouds are generated following a power-law density distribution ($p=-1$ for $\rho \propto r^{p}$) in a radial direction. 

We additionally consider a damping factor ($d\equiv {\rm SF_{\infty, MIR}}/{\rm SF_{\infty,opt}}$), which is the fraction of the variability amplitude between the optical and the MIR. This term is necessary because the variability amplitude of the MIR data is significantly smaller than that of the optical data (i.e., $d < 1$; \citealt{neugebauer_1989, Kozlowski_2010, kozlowski_2016a}). We adopt the Python codes from \citet{li_2023}\footnote{https://github.com/bwv1194/geometric-torus-variability} but modify them by adding the damping factor in order to fit the MIR SFs more robustly. The grid of predicted MIR SFs, generated with the input optical light curve and the torus transfer function computed from the torus model, spans the following range of parameters:  $-1.85 \le \log (R_{\rm in}/{\rm pc}) \le -0.1$, $0^\circ < i < 45^\circ$, $1.3 \le Y \le 1.5$, $\sigma=30^\circ$, and $0.45 \le d \le 0.8$. While the range of $R_{\rm in}$ is determined based on the results from the previous NIR and MIR RM projects \cite[e.g.,][]{koshida_2014, lyu_2019}, that of $Y$ is inferred from the morphology of hot dust derived from the NIR interferometric observations \cite[e.g.,][]{kishimoto_2009}. We use the grid of modeled SFs to fit the observed MIR SFs from \citet{son_2023b}. Because the optical SFs only securely cover $\Delta t < 1000$ days, we use the same range of $\Delta t$ to fit the MIR SFs.    

The transfer function and, therefore, the shape of the MIR SFs are mostly determined by the combination of $R_{\rm in}$, $Y$, and a power-law index for the density distribution \cite[$p$; e.g.,][]{almeyda_2020}. To keep the consistency of the method with \citet{li_2023}, we first adopt the small $Y$ and a fixed $p$. 
The best fit for the entire sample yields $R_{\rm in} = 0.18 \pm 0.01$ pc (Figure 4{\it a}), which is in good agreement with the torus sizes derived from the reverberation mapping method and interferometric observations of nearby Seyferts [$42.5 \lesssim \log~(L_{\rm bol}/{\rm erg~s^{-1}})\lesssim 46.5$] at $\sim2.2~\mu {\rm m}$ \citep[e.g.,][]{kishimoto_2011,koshida_2014}.

\begin{table*}
\caption{Radius of the torus\label{tab:table2}}
\centering
\begin{tabular}{ccccc}
\hline \hline
Subsample &
$\log (L_{\rm bol}/{\rm erg~s^{-1}}) $ &
$\log (R_{\rm in}/{\rm pc})$ &
$\log (R_{\rm eff}/{\rm pc})$ &
$\lambda_{\rm rest}$\\
(1) &
(2) &
(3) &
(4) &
(5) \\
\hline
All                           &\h $45.21\pm0.20$ &\h$-0.75\pm0.01$ &\h $-0.69\pm0.01$& \h$2.63\pm0.11$\\
$\log L_{5100} \leq 44.67$    &\h $44.51\pm0.11$ &\h$-1.01\pm0.06$ &\h $-0.92\pm0.06$& \h$2.75\pm0.10$\\
$44.67<\log L_{5100}\leq45.17$&\h $45.01\pm0.10$ &\h$-0.76\pm0.01$ &\h $-0.70\pm0.01$& \h$2.70\pm0.11$\\
$44.67<\log L_{5100}\leq45.17$&\h $45.34\pm0.10$ &\h$-0.70\pm0.01$ &\h $-0.64\pm0.01$&
\h$2.58\pm0.09$\\
$45.67<\log L_{5100}$         &\h $45.80\pm0.09$ &\h$-0.42\pm0.02$ &\h $-0.36\pm0.02$& \h$2.52\pm0.06$\\
\hline
\end{tabular}
\tablefoot{
Col. (1): Subsample, where $L_{\rm bol}$ is the AGN bolometric luminosity in units of erg s$^{-1}$.
Col. (2): Median $L_{\rm bol}$ and its median absolute deviation.
Col. (3): Inner radius of the torus.
Col. (4): Effective radius of the torus, measured from the effective $\tau$. 
Col. (5): Effective wavelength of the W1 filter in the rest-frame. 
}
\end{table*}

\begin{figure}[tbp!]
\centering
\includegraphics[width=0.45\textwidth]{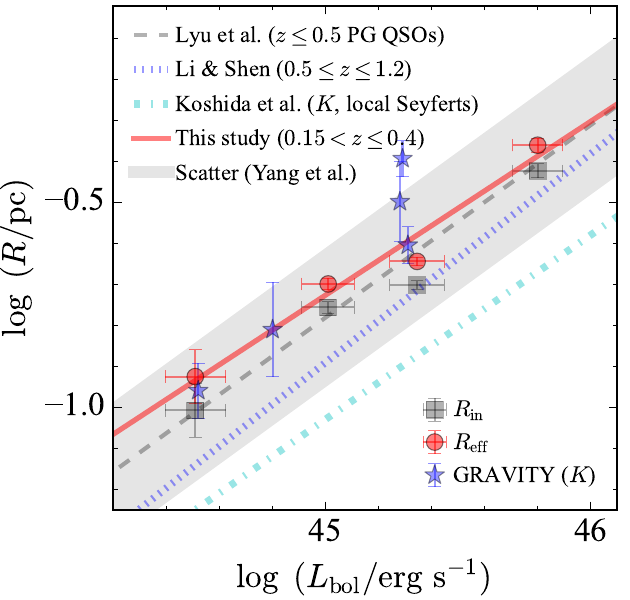}
\caption{
The size-luminosity relation of the torus of type 1 AGNs. Gray squares denote the inner radius of the torus derived from fitting the MIR SFs with the toy model for the torus. Red circles represent the effective size of the torus estimated from the best-fit transfer function, with the red solid line giving their ordinary distance regression fit to the size-luminosity relation. The uncertainties of $R_{\rm in}$ are the same as those of $R_{\rm eff}$. The grey dashed line shows the size-luminosity relation of low-redshift ($z<0.5$) PG quasars estimated using the reverberation mapping method \citep{lyu_2019}. The grey shaded area represents the $1\,\sigma$ scatter ($\sim0.17$ dex) estimated from a sample of relatively high-redshift ($0.3<z<2$) AGNs \citep{yang_2020}. The blue dotted line is the relation of intermediate-redshift ($0.5 \le z \le 1.2$) quasars in Stripe 82, based on the direct comparison between the optical SFs and MIR (W1) SFs, similar to this study. The cyan dashed-dotted line denotes the relation of the local Seyferts in the $K$ band inferred from the reverberation mapping method. Blue stars represent the size of the torus estimated from the $K$-band interferometry measurements (\citealp{gravity_2020b})}.    

\end{figure}

\section{The size-luminosity relation of the torus}
We apply the fit for the ensemble SFs to the subsamples divided by AGN luminosity. Fitting the faintest (Figure 4b) and brightest (Figure 4c) AGNs clearly shows that, as expected, $R_{\rm in}$ is correlated with the AGN luminosity (Table 2). We compare our results with those obtained from reverberation mapping studies, which investigated the size-luminosity relation of the torus using the time lag between the optical light curves obtained from ground-based telescopes and MIR light curves derived from WISE \cite[e.g.,][]{koshida_2014, lyu_2019, yang_2020}. The MIR light curves of low-redshift ($z<0.5$) Palomar-Green (PG; Schmidt \& Green 1983)  quasars studied by \citet{lyu_2019} are comparable to those of our sample. They demonstrated a strong correlation between the size of the torus determined in the W1 band and the AGN luminosity, in the form $R \propto L_{\rm AGN}^{0.47}$. While the intrinsic scatter in the size-luminosity relation was not reported in \citet{lyu_2019}, using more distant quasars at $0.3<z<2$ \citet{yang_2020} found a size-luminosity relation similar to that of \citet{lyu_2019}; this is illustrated with the grey area in Figure 5, which has an intrinsic scatter of 0.17 dex.

For a direct comparison with previous reverberation mapping studies, we compute the effective radius ($R_{\rm eff}$) of the torus determined from the best-fit transfer function. This is equivalent to the effective time delay ($\tau_{\rm eff}$) directly measured from the reverberation mapping method, which is expressed as
\begin{equation}
\tau_{\rm eff} = \frac{\int{\tau \Psi(\tau) d\tau}}{\int{\Psi({\tau}) d\tau}}.
\end{equation}
The effective time delay depends on the inner radius and the geometry of the torus (e.g., the inclination angle and the ratio of the outer and inner radius). Figure 5 shows that the size-luminosity relation from our study is in good agreement with previous studies within the intrinsic scatter. An ordinary distance regression for our measurements yields 

\begin{equation}
\begin{split}
\log~(R_{\rm eff}/{\rm pc}) =  (0.42\pm0.06)\times \log~(L_{\rm bol}/{\rm erg~s^{-1}}) \\
- (19.81\pm2.59).
\end{split}
\end{equation}

\noindent
This result demonstrates that the ensemble SF can be as effective as the reverberation mapping method in statistically estimating the size of the torus. Interestingly, the zero point of the size-luminosity relation from this study is slightly larger than that from \citet{li_2023}. The offset in the torus size ranges from 0.1 to 0.2 dex at $\log\ (L_{5100}/{\rm erg\ s^{-1}}) = 44.5-46.$ This may be attributed to the dependence of the torus size on the rest-frame wavelength \cite[e.g.,][]{lyu_2019}. The rest-frame wavelength ($\lambda_{\rm rest} \approx 1.4-2.8\, \mu{\rm m}$) of the MIR data in the sample of \citet{li_2023} is significantly shorter than that of our sample ($\lambda_{\rm rest} \approx 2.5-2.8\, \mu{\rm m}$). The shorter wavelengths likely probe the inner regions of the torus. In support of this interpretation, we note that the $K$-band time lag from \cite{koshida_2014} ($\lambda_{\rm rest} \approx 2\, \mu{\rm m}$) is marginally shorter than that inferred from the W1-based studies. Interestingly, our estimation agrees well with the results from the $K$-band interferometry measurements (\citealp{gravity_2020b}). As our results depend on the adopted model for the torus geometry, the discrepancy in the size-luminosity relation among the different studies may originate from the methodology.

The studies with the MIR interferometry showed that the MIR emitting region ($8.5-13\ \mu$m ) is more extended than the NIR emitting region \cite[e.g.,][]{kishimoto_2011}. To account for this effect, we examine the dependence of $R_{\rm eff}$ on the outer-to-inner radius ratio ($Y$) and the radial density distribution ($p$). The SF fits with a model from $1.5\leq Y \leq 10$ and $-0.5 \leq p \leq -1.0$ yield worse results compared to the best-fits with a smaller $Y$ ($1.3\leq Y\leq 1.5$) and a fixed $p=-1$. In addition, the inferred $R_{\rm eff}$ becomes remarkably smaller than that from the best-fit results with the smaller $Y$ by up to $\sim0.5$ dex. This result is not physically meaningful as the inner radius is substantially smaller than the sublimation radius. It reveals that the hot dust emitting the MIR continuum at 2.5-2.8 $\mu$m is likely to locate the innermost region of the torus and may not be extended along the radius, which is consistent with the previous studies \cite[e.g.,][]{kishimoto_2009, kishimoto_2011}.

\section{Conclusions}
We derive the optical ensemble structure function of the low-redshift and moderately luminous AGNs selected from SDSS using the $g$-band multi-epoch data obtained from the ZTF survey. We find that a broken power law, similar to the damped random walk, can well describe the ensemble SF. However, more luminous AGNs tend to have steeper slopes in the ensemble SF at short timescale than less luminous AGNs, which may be due to the geometric effect of the accretion disk. Consistent with previous studies, the variability amplitude also is anti-correlated with the AGN luminosity.    

Using the toy model of the torus geometry from \citep{li_2023}, we make a comparison between the optical and MIR SFs to estimate the effective size of the torus. As with reverberation mapping experiments, we find that the effective size of the torus positively correlates with AGN luminosity. The slope of the size-luminosity relation agrees well with that of previous studies, although the zero point of the relation is marginally higher than that derived from the analysis of $K-$band data for local AGNs and W1-band data for high-redshift AGNs, which implies that the size of the torus increases with increasing rest-frame wavelength. These results demonstrate that comparison between the optical and MIR SFs for a large sample can be a powerful tool for estimating the size of the torus.

\medskip
\begin{acknowledgements}
 We thank the anonymous referee for the constructive comments that substantially improved the manuscript.
This work was supported by the National Key R\&D Program of China (2022YFF0503401), the National Science Foundation of China (11991052, 12233001), the China Manned Space Project (CMS-CSST-2021-A04, CMS-CSST-2021-A06), and the National Research Foundation of Korea (NRF), through grants funded by the Korean government (MSIT) (Nos. 2022R1A4A3031306, 2023R1A2C1006261, and RS-2024-00347548).
\end{acknowledgements}

\medskip

\bibliography{torus}

\end{document}